\documentstyle[11pt,mrs2001,epsfig1_7a]{article}
\begin{document}
\title{WHERE THE VERY BRIGHT MATTER IS A STANDARD RULER}

\author{B.F. ROUKEMA$^{1,2,3}$, G.A. MAMON$^4$, S. BAJTLIK$^2$}
\affil{$^1$DARC/LUTH, Observatoire de Paris-Meudon, 5, place Jules Janssen, 
F-92.195, Meudon Cedex, France}
\affil{$^2$Nicolaus Copernicus Astronomical Center, 
ul. Bartycka 18, 00-716 Warsaw, Poland}
\affil{$^3$University of Warsaw, Krakowskie Przedmie\'scie 26/28, 
00-927 Warsaw, Poland}
\affil{$^4$Institut d'Astrophysique de Paris, 98bis Bd Arago, F-75.014 Paris,
France}

\newcommand\Omm{\Omega_{\mbox{\rm \small m}}}
\newcommand\hMpc{\mbox{$\,h^{-1}\,$Mpc}}

\begin{abstract}
   Where's the matter? The answer seems to be: ``distributed according
to a power spectrum with at least one feature (local maximum) near
120--130{\hMpc}''. Analyses of the Iovino, Clowes \& Shaver
quasar candidate catalogue at
$z\sim2$ and the 2dF Quasar Survey 10K Release (2QZ-10K)
support this claim, which has previously been made both for
low redshift survey analyses and for high ($z\sim3$) redshift surveys
will be presented. This feature (i) offers a comoving standard ruler 
which can lift the matter density--cosmological constant 
($\Omm$--$\Omega_\Lambda$) degeneracy and (ii) 
might be due either to baryonic acoustic oscillations or to Planck epoch
physics which survives through inflation. (i) The 95\% confidence
constraint from the 2QZ-10K is
$\Omm= 0.25\pm0.15, \Omega_\Lambda=0.60{\pm0.35}$. This constraint
is independent of cosmic microwave background constraints and type Ia 
supernovae constraints. The only assumptions required are (a) that 
the Universe satisfies a perturbed Friedmann-Lema\^{\i}tre-Robertson-Walker
model with a possibly non-zero cosmological constant, (b) that the
density perturbations in this model on large scales ($\gg 10h^{-1}$Mpc)
remain small (``linear'') and approximately spatially 
fixed in comoving coordinates, (c) that the statistics (power spectrum 
or correlation function) of the perturbations are redshift independent,
and (d) that quasar redshifts are cosmological.
\end{abstract}

``Where's the Matter?'' 

Clues from several surveys, mostly at low redshifts, a few at high
redshifts (see the full list of references in \cite{KirCh00}) and the
increasingly significant results in analyses of quasar surveys
\cite{Deng94,RM00,RM01,RMB01} imply that the answer is ``distributed
according to a power spectrum with at least one feature (local
maximum) near 120--130{\hMpc}''.

Because the radial distance-redshift relation and the curvature
relation between radial and tangential distances both depend on local
cosmological parameters $\Omm$ (the matter density) and
$\Omega_\Lambda$ (the cosmological constant), any fine structure in
the statistics of density perturbations (traced by any collapsed objects
such as galaxies, clusters of galaxies, quasars) will appear to occur
at incorrect comoving length scales if incorrect values of $\Omm$ and
$\Omega_\Lambda$ are assumed when transforming angular positions
and redshifts into comoving spatial positions. 

Fine structure features will only appear at the correct length scales
for the correct values of $\Omm$ and $\Omega_\Lambda$.

In \cite{RM00} it was found that the $z\sim 2$ quasar candidate survey of
\cite{IovCS96} had a feature at the low redshift value of
$130${\hMpc}, provided that $0.1 < \Omm < 0.45$ (68\% confidence), for
any value $0 \le \Omega_\Lambda\le 1$. 

The initial release of the 2QZ-10K \cite{Croom01} yields much 
stronger results. The 95\% confidence
constraint from the 2QZ-10K is
$\Omm= 0.25\pm0.15, \Omega_\Lambda=0.60{\pm0.35}$. In this case, an 
{\it a priori} value of the length scale for a feature was {\it not} chosen;
internal consistency of features between sub-samples 
in the different redshift ranges 
$0.6 < z < 1.1$, 
$1.1 < z < 1.6$ and
$1.6 < z < 2.2$ was the only requirement. This yielded the constraint
on $\Omm$ and $\Omega_\Lambda$ and the length scale
$2L= 244\pm17${\hMpc}, consistent with a harmonic of the scale found 
in \cite{RM00} and by many earlier authors.

\begin{figure}
\plotone{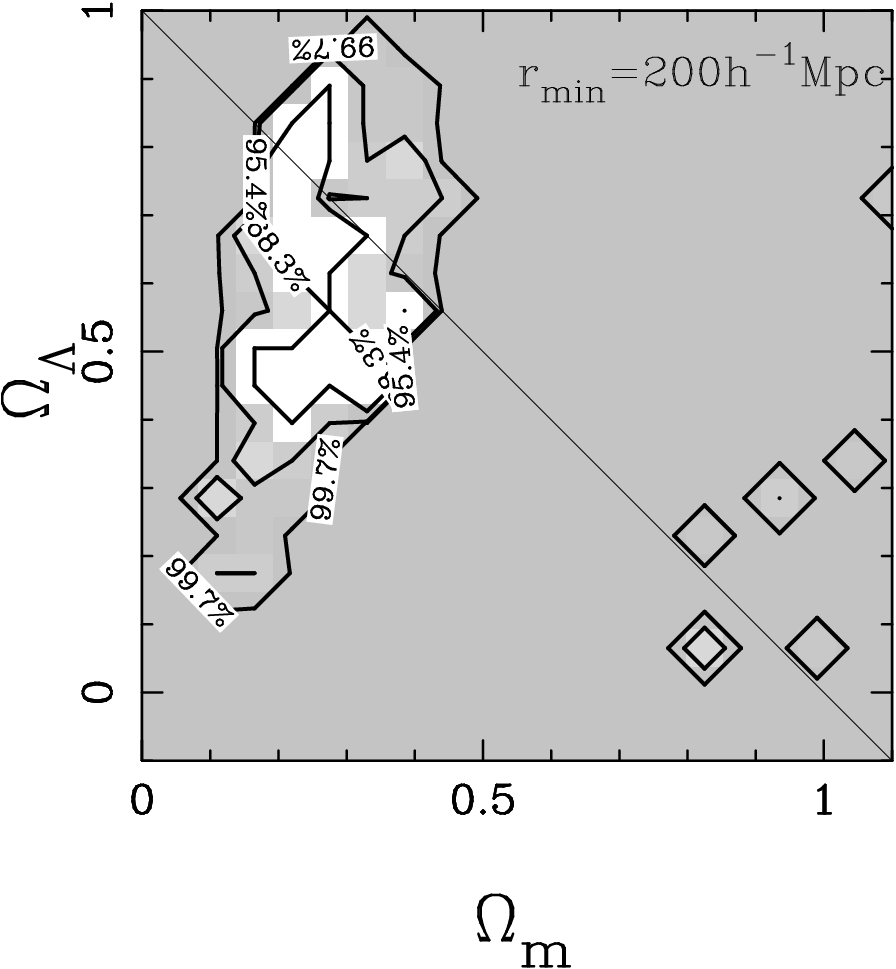}{.55\textwidth}
\caption{Confidence intervals for rejecting the hypothesis that
the first local maximum at scales greater 
than 200{\hMpc} in the comoving spatial correlation function of
the 2QZ-10K quasars occurs at the same comoving scale in all three
redshift intervals. It is clear that only one region of the
$\Omm$--$\Omega_\Lambda$ plane is favoured.}
\end{figure}

\acknowledgements{
This research has been supported by the 
Polish Council for Scientific Research Grant
KBN 2 P03D 017 19
and has benefited from 
the Programme jumelage 16 astronomie 
France/Pologne (CNRS/PAN) of the Minist\`ere de la recherche et
de la technologie (France).}

\newcommand\joref[5]{#1, #5, {#2 }{#3, } #4}  
\newcommand\epref[3]{#1, #3, #2}

\vfill
\end{document}